\documentstyle[12pt]{article}

\begin{document}
\topmargin 0pt
\oddsidemargin 7mm
\headheight 0pt
\topskip 0mm

\addtolength{\baselineskip}{0.40\baselineskip}

\hfill KAIST-CHEP-95/19

\hfill December 1995

\begin{center}

\vspace{36pt}
{\large \bf Remarks on the Upper Bounds on Higgs Boson Mass from Triviality}

\end{center}

\vspace{36pt}

\begin{center}

Jae Sik Lee
\footnote{e-mail address: jslee@chep6.kaist.ac.kr} and Jae Kwan Kim

\vspace{20pt}

{\it Department of Physics,  \\
Korea Advanced Institute of Science and Technology, \\
Taejon 305-701, Korea } \\

\end{center}

\vspace{10pt}

\vfill

\begin{center}
{\bf ABSTRACT}
\end{center}
We study the effects of the one-loop matching conditions on Higgs boson
and top quark masses 
on the triviality bounds on the Higgs boson mass using
$\beta_{\lambda}$ with corrected two-loop coefficients.
We obtain quite higher results than previous ones and observe
that the triviality bounds are not nearly influenced by varying top quark mass 
over the range measured at CDF and D0.
The effects of typo errors in
$\beta_{\lambda}^{(2)}$ and the one-loop matching condition on the top quark
mass are negligible. 
We estimate the size of effects on the triviality bounds from the one-loop
matching condition on the Higgs boson mass.

\vspace{12pt}

\begin{flushleft}
\end{flushleft}

\noindent

\vspace{24pt}

\vfill

\vspace{15pt}

\newpage

The tree level Higgs potential in the standard model (SM) is
given by
\begin{equation}
V=\frac{1}{2}m^2\phi^2+\frac{\lambda}{24}\phi^4.
\end{equation}
As well known, a pure scalar $\lambda \phi^4$ theory allows only
$\lambda_{\rm R}=0$ where $\lambda_{\rm R}$ is the renormalization
coupling.  If $\lambda_{\rm R}$ is not zero there is a singularity
in the evolution of the running coupling constant.
This is the triviality problem of the pure $\lambda \phi^4$
theory.

But one can expect that the interactions of the standard Higgs with other
particles in the SM can make $\lambda_{\rm R}$ to have a
non-zero value. Beg et al. [1] studied this possibility with a consideration
of the fact that the $U(1)$ coupling $g_1$ always produces a Landau
singularity at a very high scale $\Lambda_{\rm LS} \approx 10^{42}$ GeV.
They obtained the upper bound of the Higgs boson mass as a function of the top 
quark mass from a condition of $\Lambda_{\lambda} > \Lambda_{\rm LS}$ where
$\Lambda_{\lambda}$ is  a scale of the singularity of the running coupling
$\lambda$. (actually they used $\Lambda_{\rm B}=4 \times 10^{37}$ GeV
instead of $\Lambda_{\rm LS}$)

But many people think that the SM is a low energy effective theory
which is embedded in some more fundamental theory at a scale $\Lambda$.
Usually this scale $\Lambda$ is less than $\Lambda_{\rm LS}$.
For example, the Planck scale $\Lambda_{\rm PL} = 10^{19}$ GeV where
something new must appear
is mush less than $\Lambda_{\rm LS}$. Therefore it was required to calculate
the triviality bounds on the Higgs boson mass in the case that the new
physics scale $\Lambda$ is less than $\Lambda_{\rm LS}$. Lindner et al. 
implemented this calculation at one- and two-loop level. [2,3]

Grzadkowski and Lindner implemented, at two-loop level, analyses of the
triviality Higgs boson mass bounds in the SM using tree level matching 
conditions on Higgs boson and top quark masses. [3] And they expect 
the effects of the one-loop matching
condition on Higgs boson mass to be small for $M_H < 500$ GeV.

It was noted that there were two errors in the $\beta$ function
of $\lambda$,$~~\beta_{\lambda}$. [4] These errors are the electroweak 
contributions to the two-loop coefficients of the $\beta_{\lambda}$ 
function. [5] We denote the two-loop part of the $\beta_{\lambda}$ function
as $\beta_{\lambda}^{(2)}$.

Considering only the interactions of gauge particles and top quark with
the Higgs particle, the one-loop matching conditions are
\begin{eqnarray}
\lambda(\mu_0) &=& 3 \frac{M_H^2}{v^2}\left[1+\delta_{\lambda}(\mu_0)\right],
\nonumber \\
h_t(\mu_0) &=& \sqrt{2} \frac{M_t}{v}\left[1+\delta_t(\mu_0)\right],
\end{eqnarray}
where $h_t$ is top-Yukawa coupling, $\mu_0$ is the renormalization
scale and $v=(\sqrt{2} G_F)^{-1/2}=246$ GeV. $M_H$ and
${M_t}$ denote Higgs boson and top quark mass respectively. Explicit form of 
$\delta_{\lambda}(\mu_0)$ and $\delta_t(\mu_0)$ can be found in Ref. [6].

Recently the existence of the top quark is clear. [7]
\begin{equation}
M_t =180 \pm 12 {\rm GeV}~~~~{\rm (CDF+D0)}.
\end{equation}
For this range of $M_t$ value and $M_H < 1$ TeV, $\delta_t(\mu_0)$ is 
$\sim 5$ \% or less for $\mu_0 = M_Z$ and $M_t$. So this effect is negligible
in the study of the upper bounds on the Higgs boson mass. But $\delta_{\lambda}(\mu_0)$
is quite large and heavily depends on the choice of $\mu_0$.

In this paper we study the effects of matching condition on the Higgs boson mass 
on the triviality bounds on the Higgs boson mass using
$\beta_{\lambda}$ with corrected two-loop coefficients. 

As noted by Grzadkowski and Lindner, the evolution
behaviors of $\lambda$ is drastically changed at two-loop level. [3] 
At two-loop level
there is no more singularity, there exists a new UV fixed-point.
As scale increases, two-loop part 
of $\beta_{\lambda}$, $~\beta_{\lambda}^{(2)}$, starts to dominate and
exactly cancels the one-loop contribution at a very high scale.
This is a signal of a breakdown of a perturbation expansion. 
So the existence of this UV fixed-point doesn't mean that one can remove
the singularity without embedding since the point takes place behind 
a perturbation area.
They study the dependence
of $\lambda(\mu=0)$ as a function of $\lambda(\mu=\Lambda)$ and find a plateau
in the plane ($\lambda(0)$, $\lambda(\Lambda)$). This plateau originates from
the existence of the UV fixed-point. So the existence of the plateau is a
signal of a breakdown of a perturbation expansion and corresponds to the pole. 
From their automatic fixing procedure designed to find the end of the plateau,
they obtained the value of ${\bar \lambda}(\Lambda)$ which is about 60
\footnote{Note that the coefficient of the quartic term of the scalar potential 
which we are using is not $\frac{\lambda}{8}$ but $\frac{\lambda}{24}$.
We are using the form of the potential given in Ref. [4]. 
The Higgs potential used by authors of Ref. [2,3] corresponds to
$V=\frac{1}{2}m^2\phi^2+\frac{\lambda}{8}\phi^4$. Therefore the value of our
$\lambda$ is three times larger than that in Ref. [2,3]} 
for $M_t < 210$ GeV and $\Lambda < 10^{15}$ GeV.

For numerical calculations we give boundary conditions for the gauge
couplings at $M_Z$ as follows
\begin{eqnarray}
g_1(M_Z) &=& 0.3578, \nonumber \\
g_2(M_Z) &=& 0.6502, \nonumber \\
\alpha_s(M_Z) &=& \frac{g_3^2(M_Z)}{4 \pi} = 0.123.
\end{eqnarray}
We use $\beta$-functions found in Ref. [4].

In Fig. 1 we show the running of $\lambda$ for $M_t = 180$ GeV and
$M_H = 200,~~300,~~400,~~500$ GeV. We use tree level matching condition with
$\mu_0=M_Z$. For each value of $M_H$, diverging line is
the result from one-loop calculations and the other is that from two-loop ones.
We observe that the effects of the corrected coefficients of $\beta_{\lambda}^{(2)}$
on the running of $\lambda$ is negligible for $M_H > 200$ GeV. 
So the effects of wrong coefficients of $\beta_{\lambda}^{(2)}$ on bounds on the Higgs boson
mass are negligible and the results of Ref. [3] are not influenced
by these corrections. From Fig. 1 we can
observe the existence of the UV fixed-point where $\lambda$ is about 72.
The scale where one-loop calculation diverges and the other scale where
two-loop calculation starts to be stationary are close to each other. 
And we can see the effects of the two-loop corrections diminishes the slope of
$\lambda(\mu)$. This results from the fact that $\beta_{\lambda}^{(2)}$ 
is negative. The corresponding pole position at two-loop level is expected
to be smaller than that at one-loop level.
Therefore we can expect that triviality bounds computed at two-loop
level are higher than those computed at one-loop level. 
When we calculate the corresponding pole position at two-loop level we
require the calculation is consistent with this expectation.

We define $\Lambda_p^{(1)}$ as the pole position of $\lambda$ computed at 
one-loop level using tree level matching condition with $\mu_0=M_Z$.
In Fig. 2 we plot $\Lambda_p^{(1)}$ as a function of $M_H$ for three values of
$M_t=160,~180,~200$ GeV. This is the very same framework as that of Ref. [2].
A requirement $\Lambda_p^{(1)} > \Lambda$ gives the upper bound on $M_H$ as a function of
$\Lambda$. [2] This requirement means that the triviality is removed by adding
fields which couple to the Higgs particle in the new physics.
Numerically, we identify the pole position as the scale 
at which $\lambda$ starts to be bigger than 1000.
If news physics appears at the Planck scale, $M_H < \sim 220$
GeV to satisfy the requirement $\Lambda_p^{(1)} > \Lambda_{\rm PL}$. This result is different 
from that of Ref. [2]. Our result is somewhat higher than that of Ref. [2].
From Fig. 2 we observe that the effects of varying $M_t$ from 160 to 200 GeV 
on the Higgs boson mass bound is less than about 10 GeV when 
$\Lambda <10^{20}$ GeV. Contrary to the results of Ref. [2] 
we can not observe the large effects of varying $M_t$.
It seems like that the origin of these discrepancies come from the differences
of numerical treatments of running of $\lambda$ and the singularity.
But the origin is unclear. We observe that the smaller $\Lambda$, 
the less dependence on $M_t$.

To study the effects of two-loop corrections to $\beta$ functions and one-loop 
matching condition on the Higgs boson mass bounds, we consider following
two cases.
The first one is the case of tree level matching condition with $\mu_0=M_Z$ 
and two-loop $\beta$ function. The second one is the case of one-loop matching 
condition with $\mu_0=M_Z,~M_t,~M_H$ and one-loop $\beta$ function. 
The corresponding pole positions of the first and second cases are denoted by
$\Lambda_p^{(2)}$ and $\Lambda_p^{(1),(\mu_0)}$ respectively.

In Fig. 3 we plot $\Lambda_p^{(2)}$ as a function of $M_H$ for $M_t=180$ GeV
where $\Lambda_p^{(2)}$ is the scale of perturbative expansion breakdown at
two-loop level using tree level matching condition.
We also plot $\Lambda_p^{(1)}$ for comparisons.
$\Lambda_p^{(2)}$ is defined as the scale at which $\lambda$ starts to be bigger 
than some value $\lambda^{\rm cut}$ around the scale $\Lambda_p^{(1)}$, 
or $\lambda\left(\mu > \Lambda_p^{(2)}\right) > \lambda^{\rm cut}$. 
We choose three values $\lambda^{\rm cut} = 40,~60,~70$ to examine the effects
of choice of the value of $\lambda^{\rm cut}$. 
We take $\lambda^{\rm cut}=60$ because
this value is the one used in Ref. [3] in interested region of $M_t$.
We take $\lambda^{\rm cut}=70$ because this value is very near to the 
UV fixed-point. We observe that the case of $\lambda^{\rm cut} = 40$ 
produces lower bound than one-loop case for $\Lambda < 10^5$ GeV.
Although smaller value of $\lambda^{\rm cut}$ is more realistic 
in view of the demand that the two-loop effects make sense preturbatively,
the choice of small $\lambda^{\rm cut}$ produces lower bound than that
obtained by calculations at one-loop level. This is not consistent with 
the general expectation of the two-loop effects. We expect that
two-loop effects on bound are positive since $\beta_{\lambda}^{(2)}$
is negative. Therefore we estimate an uncertainty of the two-loop effects
by varying $\lambda^{\rm cut}$ from 40 to 70.
Taking 60 as a center value and for $M_t=180$ GeV, 
we estimate the bounds on the Higgs boson mass as follows
\begin{eqnarray}
M_H &<& 270~\pm 10~{\rm GeV}~~~{\rm for}~~\Lambda=10^{15} {\rm GeV} \nonumber \\
M_H &<& 340~\pm 25~{\rm GeV}~~~{\rm for}~~\Lambda=10^{10} {\rm GeV} \nonumber \\
M_H &<& 500~\pm 70~{\rm GeV}~~~{\rm for}~~\Lambda=10^{6} {\rm GeV}. \nonumber 
\end{eqnarray}
The triviality bound has a tendency to increase when we consider 
two-loop effects and these increments are
larger for smaller $\Lambda$. Since these theoretical errors are quite
large, the effects of the matching condition are not important if the
effects are small.

We denote the pole position computed at one-loop level using the 
one-loop matching condition on the Higgs boson mass as $\Lambda_p^{(1),(\mu_0)}$. 
In Fig. 4 we plot $\Lambda_p^{(1),(\mu_0)}$ as a function of $M_H$ for three choices
of the renormalization scale $\mu_0=M_Z,~M_t,~M_H$ and $M_t=180$ GeV. One can find
a rather similar figure in Ref. [8] From this figure we can estimate 
the size of the effects of the one-loop matching condition
on the Higgs boson mass. For large value of $M_H$ and 
$\mu_0=M_Z,~M_t$, $\delta_{\lambda}(\mu_0)$ can be less than -1.
The fact $\delta_{\lambda}(\mu_0=M_Z,M_t)$ can be less than -1
illustrates that the perturbation results are not reliable anymore 
and $\lambda$ at those $\mu_0$ is negative. So
we compute $\Lambda_p^{(1),(\mu_0)}$ only for the values of $M_H$ satisfying
$\delta_{\lambda}(\mu_0) > -1$. $\Lambda_p^{(1),(\mu_0)}$ 
for large $M_H$ with small ($1+\delta_{\lambda}(\mu_0)$) is the same order of magnitude
as that for small $M_H$ with small $\delta_{\lambda}(\mu_0)$. From Fig. 4 we
observe that $\mu_0=M_H$ choice gives nearly the same, but a litter bit lower,
triviality bound. 
For the $M_H < \sim 500$ GeV where the perturbative results are
reliable when $\mu_0 = M_Z,~M_t$, the effects of taking into
account matching condition on Higgs boson mass give the same order of magnitude as
given by the differences between triviality bounds computed at one- or 
two-loop levels. In the case $\mu_0 = M_H$, the effects are negligible.

As a summary, we plot the triviality bound on $M_H$ as a function of $M_t$
for several $\Lambda$ values computed at two-loop level with $\lambda^{\rm cut}=60$
using one-loop matching condition with $\mu_0=M_H$. (see Fig. 5) The effects of
the one-loop matching condition with $\mu_0=M_H$ lower the bounds obtained by calculation
at two-loop level using tree level matching condition. The bounds are lowered by
10, 15 and 40 GeV for $\Lambda=10^{15}$, 10$^{10}$ and 10$^6$ GeV respectively.
We summarize as follows
\begin{eqnarray}
M_H &<& 260~\pm 10~\pm 2~{\rm GeV}~~~{\rm for}~~\Lambda=10^{15} {\rm GeV} \nonumber \\
M_H &<& 325~\pm 25~\pm 2~{\rm GeV}~~~{\rm for}~~\Lambda=10^{10} {\rm GeV} \nonumber \\
M_H &<& 460~\pm 70~\pm 7~{\rm GeV}~~~{\rm for}~~\Lambda=10^{6} {\rm GeV}. \nonumber 
\end{eqnarray}
The first and second errors are related with the choice of the value $\lambda^{\rm cut}$ and
varying top quark mass from 150 to 210 GeV respectively.
We observe that the dependence on $M_t$ are small and obtain quite higher results than those
of Ref. [2]. The theoretical errors of these bounds are larger for smaller
$\Lambda$. The effects of typo errors in
$\beta_{\lambda}^{(2)}$ and matching condition on the top quark mass on
triviality bounds are negligible.
%
\section*{Acknowledgments}
This work was supported in part by Korea Science and Engineering Foundation.

\newpage


\newpage

\section*{Figure Captions}

\begin{description}
\item{Fig. 1} : Plots of $\lambda(\mu)$ for $M_t = 180$ GeV and
$M_H = 200$ (solid line), 300 (dashed line), 400 (dotted line) and 500 (dash-dotted line) 
GeV. For each value of $M_H$, diverging line is
the result from one-loop calculations and the other is that from two-loop ones.
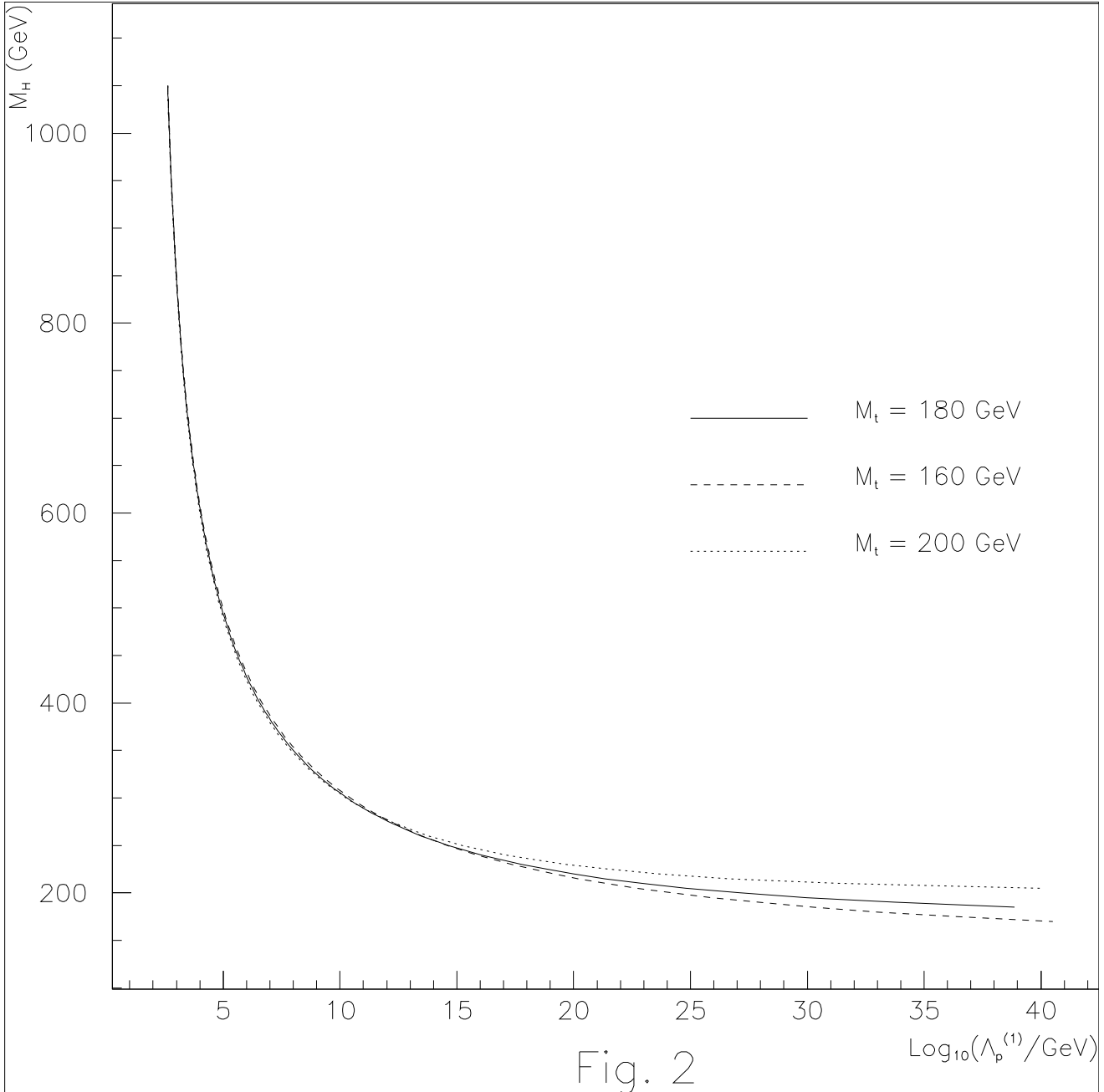
\item{Fig. 2} : Plots of $\Lambda_p^{(1)}$ as a function of $M_H$ for three values of
$M_t=160$ (dashed line), 180 (solid line) and 200 (dotted line) GeV. 
$\Lambda_p^{(1)}$ is the pole
position of $\lambda$ computed at one-loop level without using matching condition.
\item{Fig. 3} : Plots $\Lambda_p^{(1),(2)}$ as a function of $M_H$ for $M_t=180$ GeV.
$\Lambda_p^{(2)}$ is defined as the scale at which $\lambda$ starts to be bigger 
than some value $\lambda^{\rm cut}$. We denote $\Lambda_p^{(1)}$ as a solid line,
$\Lambda_p^{(2)}$ with $\lambda^{\rm cut}=$ 60 as a dashed line,
$\Lambda_p^{(2)}$ with $\lambda^{\rm cut}=$ 40 as a dotted line and
$\Lambda_p^{(2)}$ with $\lambda^{\rm cut}=$ 70 as a dashed-dotted line.
\item{Fig. 4} : Plot of $\Lambda_p^{(1)}$ (solid line) and
plots of $\Lambda_p^{(1),(\mu_0)}$ as a function of $M_H$ for three choices
of the renormalization scale $\mu_0=M_Z$ (dashed line), $M_t$ (dotted line) and $M_H$
(dash-dotted line) and $M_t=180$ GeV. We denote
the pole position computed at one-loop level using matching condition on the Higgs boson mass as $\Lambda_p^{(1),(\mu_0)}$. 
\item{Fig. 5} : Plots of the triviality bounds on $M_H$ as a function of $M_t$
for several values of $\Lambda=10^{19}$ (thick solid line), $10^{15}$ (solid line),
$10^{10}$ (dashed line), $10^{6}$ (dotted line) and $10^{3}$ (dash-dotted line) GeV.
We use two-loop $\beta$ functions with $\lambda^{\rm cut}=60$ and one-loop matching 
condition with $\mu_0=M_H$.
\end{description}

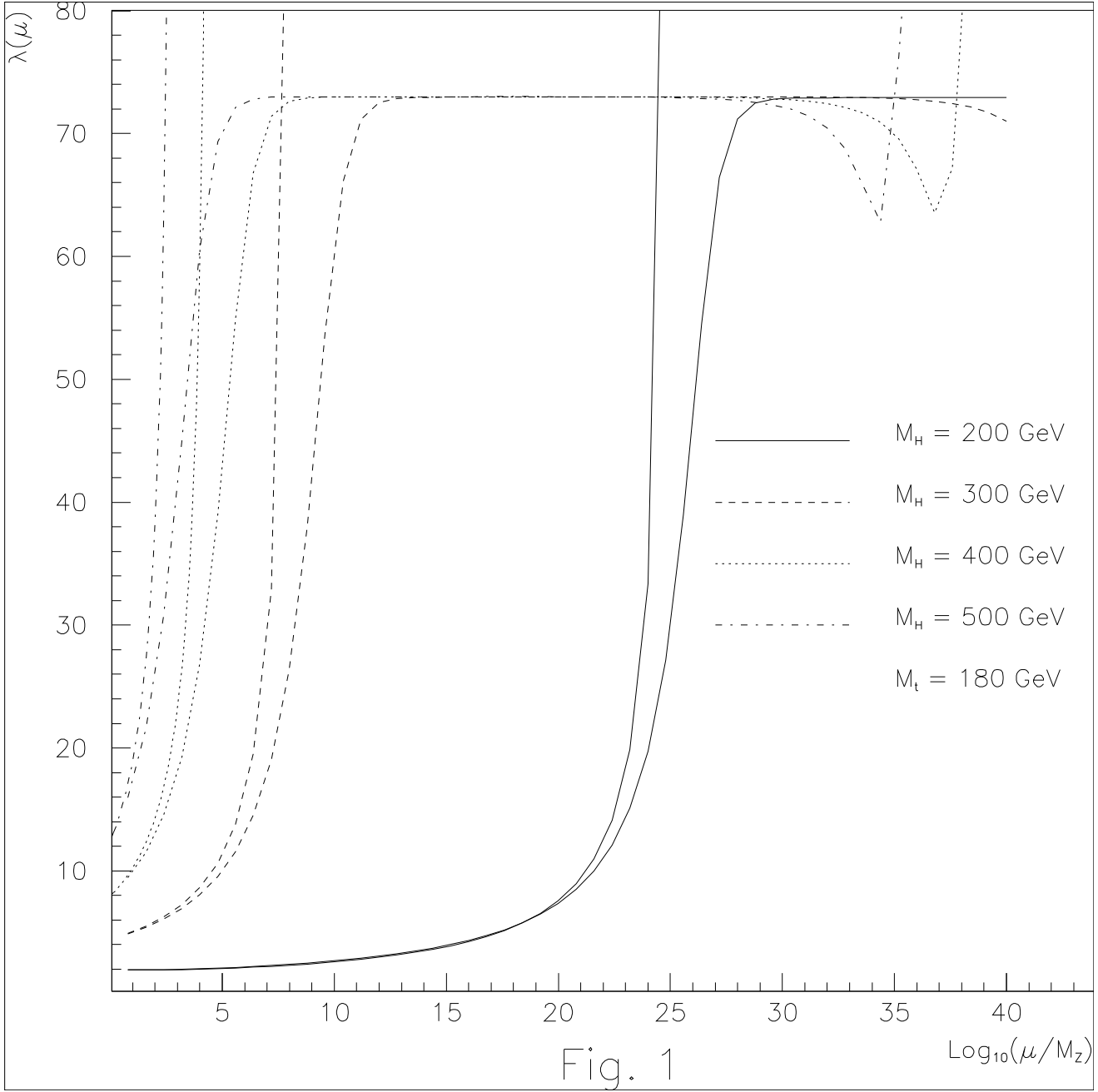
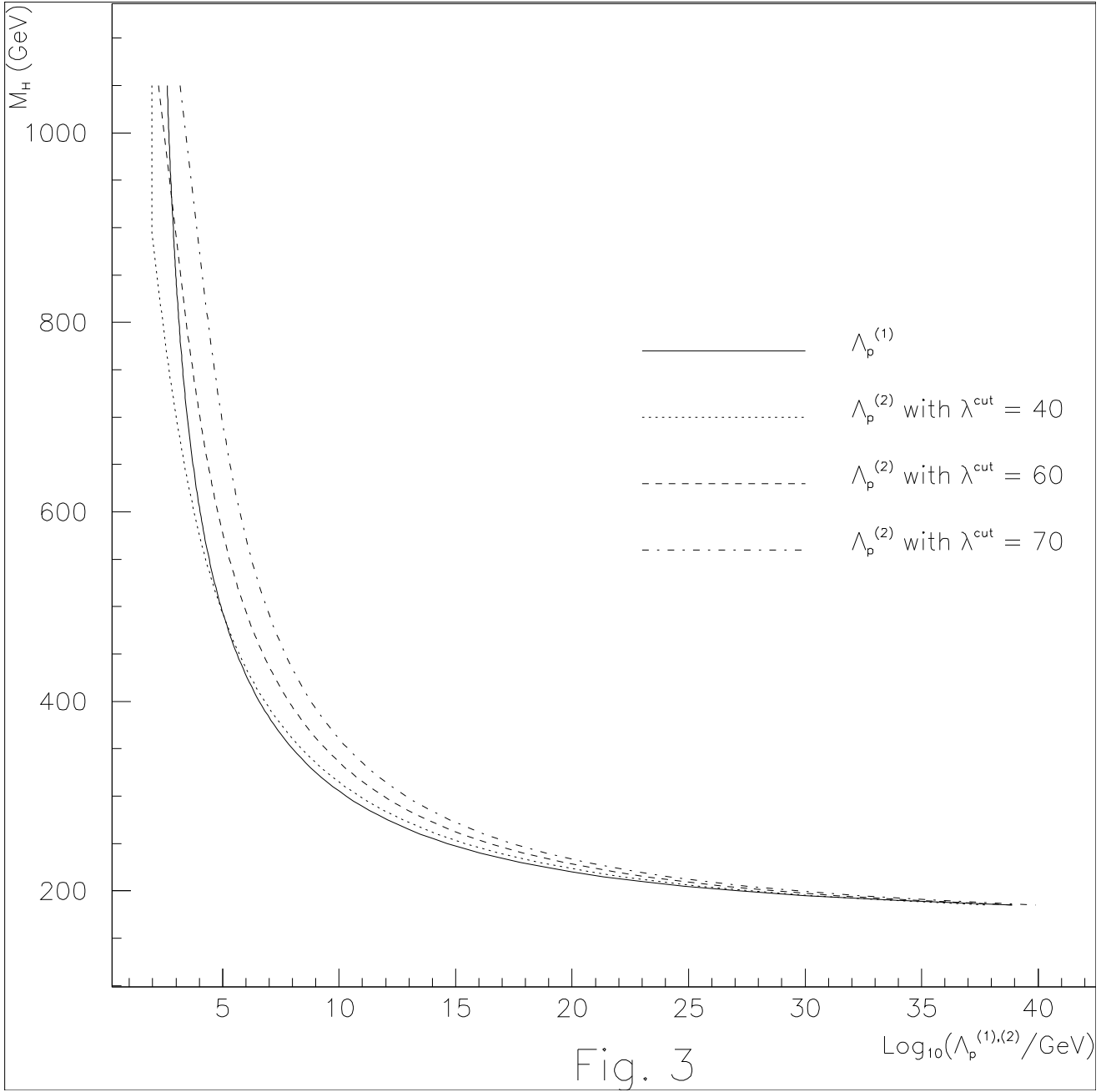
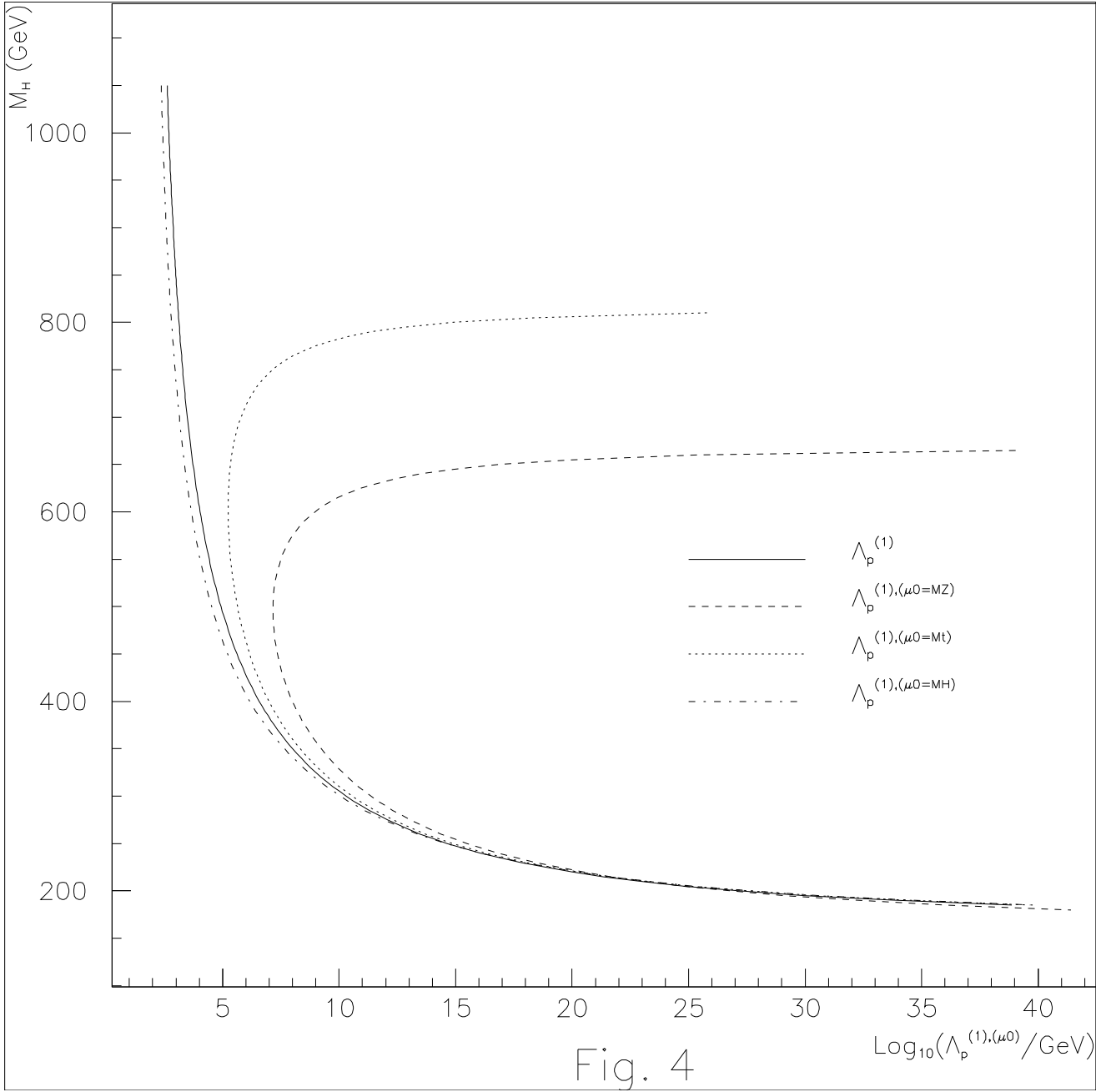
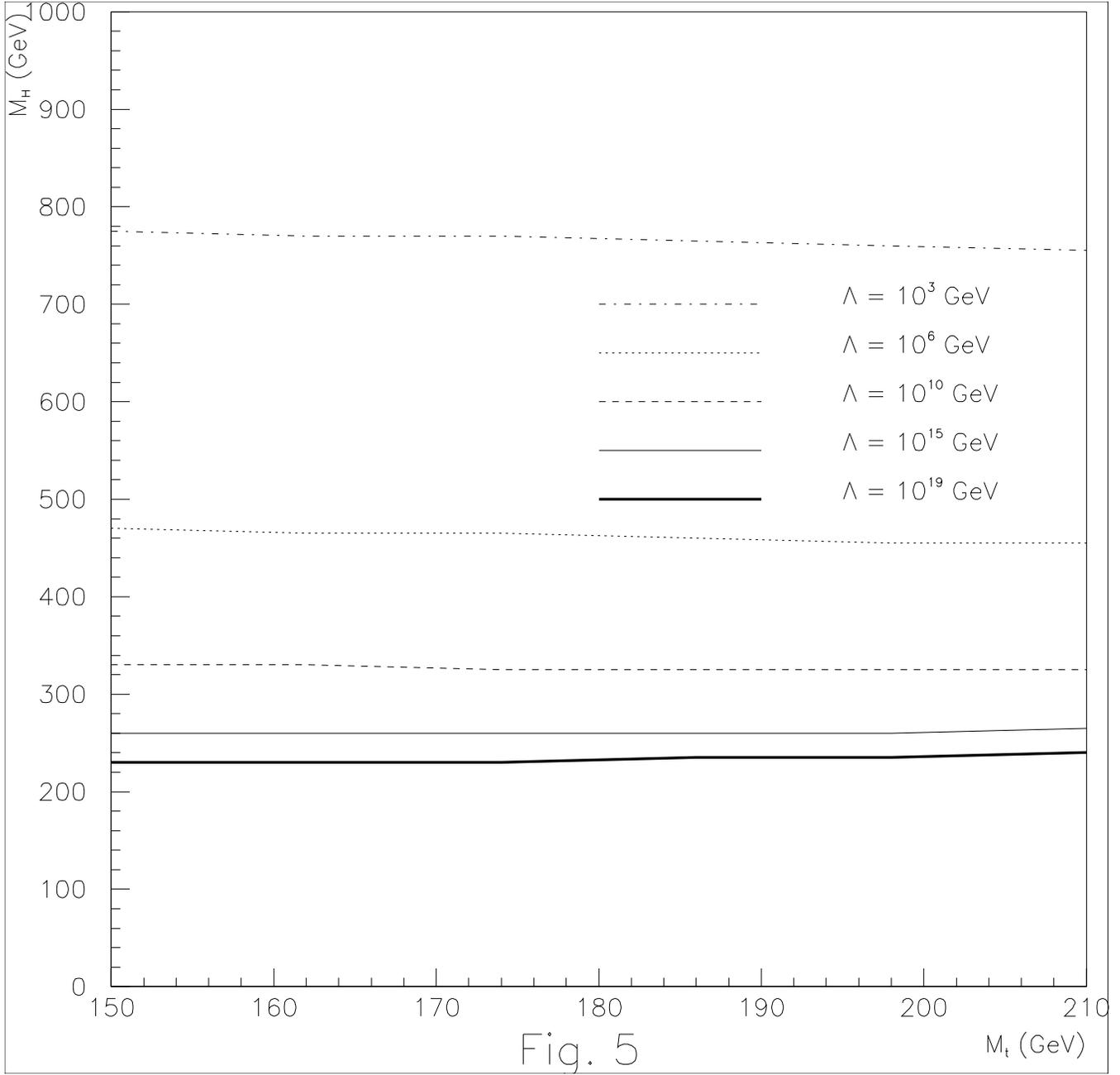
\end{document}